\documentclass[prb,floatfix,twocolumn,amsmath,amssymb,showpacs]{revtex4}
\usepackage{graphicx}

\begin{document}

\title{Magnetic and transport properties of a one dimensional 
frustrated t-J model for vanadate nanotubes}

\author{S. Costamagna and J. A. Riera}
\affiliation{
Instituto de F\'{\i}sica Rosario, Consejo Nacional de
Investigaciones Cient\'{\i}ficas y T\'ecnicas,\\
Universidad Nacional de Rosario, Rosario, Argentina}

\date{\today}

\begin{abstract}
We propose a one-dimensional model consisting of a chain with a
t-J Hamiltonian coupled to a Heisenberg chain in a frustrated
geometry to describe the appearance of the ferromagnetic phase
which has been experimentally observed in vanadate nanotubes.
This model contains a mechanism of frustration suppressed by
doping suggested by L. Krusin-Elbaum,
{\it et al.} [Nature {\bf 431}, 672 (2004)].
We study, using numerical techniques in small clusters, the relation
between magnetic order and transport properties in the proposed model,
and we perform a detailed  comparison of the physical properties of
this model with those of the ferromagnetic Kondo lattice model. For
this comparison, a number of results for the latter model, obtained
using the same numerical techniques, will be provided to complement
those results already available in the literature. We conclude that
it does not appear to be a true ferromagnetic order in the
proposed model, but rather an incommensurate ferrimagnetic one, and 
contrary to what
happens in the ferromagnetic Kondo lattice model, electronic
transport is somewhat suppressed by this ferrimagnetic order.
\end{abstract}


\maketitle

\section{Introduction}
\label{intro}

The possibility of taking advantage of the spin of the electrons in
addition to their charge in order to develop devices with new
capabilities, which is exploited in the emergent field of
spintronics\cite{wolf}, has in turn given impulse to the search for
new materials with metallic ferromagnetic (FM) phases.

An extensively studied family of compounds which presents these
properties is the family of the so called ``manganites", for 
example La$_{1-x}$Ca$_x$MnO$_3$.\cite{lamno} This family of manganese 
oxides has recently received additional interest for its property of
``colossal" magnetoresistance.\cite{colossal}

Taking into account the vast diversity of physical behavior found in
transition metal oxides, it seems natural to look for this kind of
properties within these materials. Following this program, it was
found that a vanadium oxide (VO$_x$) presents upon doping a 
ferromagnetic phase at room temperature.\cite{vanadato_tube}
This compound, as well as manganites,\cite{manganato_tube} has been
synthesized as very thin cylinders or nanowires thus opening the way to
their use as ferromagnetic leads in nanoscopic spin 
valves\cite{pasupathy,martinek} among other devices. In this compound,
double-layered nanotube ``walls" of vanadium oxide are separated by
dodecylammonium chains.\cite{vanadato_tube,votube2,votube0,votube1}
This structure is reminiscent of the carbon nanotubes which are the
essential components in important devices that have already been
developed.\cite{carbo_nano}

In manganites, it is now widely accepted that most of their relevant 
physical properties, particularly the property of magnetoresistance,
can be explained by a single-orbital ferromagnetic 
Kondo lattice (FKL) model,\cite{rierahallberg,dagotto} which can in 
turn lead to the well-known ``double exchange" (DE) model in the
limit of infinitely large Hund coupling. Of
course, it is necessary to resort to at least a two-orbital model
to describe properties related to orbital ordering.\cite{two-orbs}

The purpose of the present work is to examine an alternative 
mechanism to the one contained in the FKL or DE models. This mechanism
was suggested in Ref.~\onlinecite{vanadato_tube} and is essentially
based in the suppression of frustrated antiferromagnetic couplings
in VO$_x$ walls
by electron or hole doping. In order to study this mechanism we 
propose a model inspired by the VO$_x$ compound but we would like to 
emphasize that this model is {\em not}
aimed at a faithful description of this material.

Our model, shown in Fig.~\ref{fig1} below, is a simplified
one-dimensional (1D) version of the VO$_x$
structure.\cite{votube2,votube0,votube1} 
This model differs from a realistic
model for VO$_x$ not only in its dimensionality but also in a number
of details such as the zig-zag geometry of the chains, the alternation 
between V$_1$ and V$_2$ along the chain, and a possible dimerization
which may explain the spin gap experimentally observed in this
material. Nevertheless, we believe this model will capture the 
essential features of the mechanism suggested in
Ref.~\onlinecite{vanadato_tube} and at least qualitatively some
properties of the real materials. In fact, V$_1$-V$_2$ chains are 
structures present in VO$_x$ although coupled between them in a 
complex way. In support of the present approach, we would like to
recall that results for the one-dimensional FKL
model\cite{dagotto} have reproduced many essential properties
observed in manganites.

In the present manuscript we will mainly address, using numerical
techniques in small clusters, the relation between magnetic order and
transport properties in the proposed model, and we will perform a
detailed  comparison of the physical properties of the present model
with those of the FKL model. For this comparison, a number of results
for the latter model, obtained with the same numerical techniques,
will be provided to complement those results already available in
the literature.\cite{dagotto,garcia2,kienert}

The paper is organized as follows. In Section \ref{model} we present
the model studied and we provide details of the numerical
techniques employed. In Section \ref{magnetic} we show the
existence of various magnetic phases in the model studied. Transport
properties, in particular optical conductivity, are examined in
Section \ref{transport}. Section \ref{conclusions} is dedicated
to emphasize differences between the present model and the FKL model.

\section{Model and methods}
\label{model}

\begin{figure}
\includegraphics[width=0.42\textwidth]{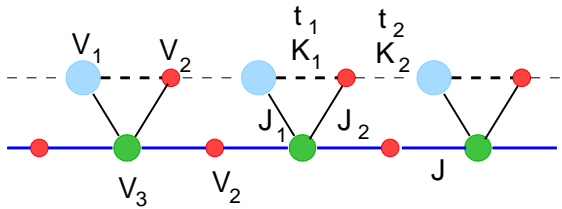}
\caption{A 1D model for VO$_x$. The upper chain, with V$_1$ and V$_2$
ions, is described by a $t-J$ model, while the lower chain, with V$_2$
and V$_3$ ions, is described by an AF Heisenberg model.}
\label{fig1}
\end{figure}

The ingredients of our effective one-band model for vanadium
orbitals only, which are common to VO$_x$, are the
following (see Fig.~\ref{fig1}). In the first place, the sites of
the system correspond to three types of vanadium ions, V$_1$,
V$_2$, and V$_3$, with a hierarchy of hole on-site energies,
$\epsilon_1 < \epsilon_2 < \epsilon_3$.\cite{vanadato_tube} V$_1$
ions are then the most likely to be occupied by holes upon doping. 

The couplings between vanadium ions are mediated by oxygens and the
relative values are determined by the quasi-pyramidal (V$_1$, V$_2$)
and tetrahedral (V$_3$) coordination between vanadium and oxygens.
In the real VO$_x$, V$_1$ and V$_2$ ions form zig-zag chains running
along a given direction in one layer and in the perpendicular 
direction in the nearest neighbor layers. These layers of chains are
connected through V$_3$ ions which are located between them. The
couplings between V$_1$, V$_2$ and V$_3$ form a frustrated
triangle of antiferromagnetic (AF) interactions.

In order to simplify the calculations we assume that double-occupancy
is forbidden, which is reasonable for d-orbitals in transition
metals, particularly close to half-filling. Then, the model for 
V-ions, with the geometry of Fig.~\ref{fig1} is defined by the
Hamiltonian:
\begin{eqnarray}
{\cal H} = &-& \sum_{i,\sigma} t_{i,i+1} ({\tilde c}^{\dagger}_{i+1 \sigma}
{\tilde c}_{i \sigma} + H.c. ) + \sum_{i} K_{i,i+1} {\bf S}_{i} \cdot
{\bf S}_{i+1} \nonumber \\
&+& J \sum_{l} ({\bf S}_{s,2l-1}\cdot {\bf S}_{s,2l} +
  {\bf S}_{s,2l}\cdot {\bf S}_{s,2l+1}) \nonumber\\
&+&  \sum_{l} (J_1 {\bf S}_{s,2l-1}\cdot {\bf S}_{2l-1} +
J_2 {\bf S}_{s,2l-1}\cdot {\bf S}_{2l})   \nonumber\\
&+& \sum_{i} \epsilon_i (1-n_{i})
\label{hamilt}
\end{eqnarray}
\noindent
where ${\tilde c}^{\dagger}_{i \sigma}$ and ${\tilde c}_{i \sigma}$
are electron creation and annihilation operators with the constraint
of no double occupancy on V$_1$ and V$_2$ sites, and ${\bf S}_j$
(${\bf S}_{s,j}$) are the spin-$1/2$ operators on the t-J (Heisenberg)
chain. The first two terms correspond to the $t-J$ Hamiltonian for
the upper chain in Fig.~\ref{fig1}, the third term to the Heisenberg
Hamiltonian for the lower chain with odd (even) numbered sites 
corresponding to V$_3$ (V$_2$) sites, and the forth term to the
exchange couplings between these two chains. For simplicity, in the
present study we will not consider dimerization in the model, and
hence we adopt the
hopping integrals $t_1 = t_2 = t$ and the exchange couplings
$K_1 = K_2 =K$, and $J_1 = J_2$. Besides, from the structure of
VO$_x$, it should be $J = J_2$ as well. All exchange couplings, $K$
and $J$, are assumed antiferromagnetic. $K=1$ is adopted as the scale
of energies. $\epsilon_2=0$ gives the reference for the on-site
energies; $\epsilon_1 <0$, which will be a variable parameter in our
study, and
$\epsilon_3=\infty$, implying a forbidden hole occupancy on V$_3$
sites. The $\epsilon_3=\infty$ condition would cause localization of
holes eventually occupying  V$_2$ sites on the lower chain in
Fig.~\ref{fig1} and then it would lead to an effective cut of this
chain. This is the reason why hole occupancy on V$_2$ sites on the
lower chain is forbidden in our model. In VO$_x$, these V$_2$ sites
would correspond to  V$_1$-V$_2$ chains running in a perpendicular
direction of the plane of Fig.~\ref{fig1}.

At half-filling, i.e. when all sites are single occupied, the system
is reduced to a frustrated spin system. In the proposed model, which
could be termed a frustrated t-J (FTJ) model, charge carriers are 
introduced by doping with holes the $t-J$ chain. Most of the results
reported below correspond to the doping fraction $x=0.4$, where
$x$ is defined as the number of doped holes divided by the number of
sites of each chain, $L$. Although most holes would go to V$_1$
sites in VO$_x$, in our model there would be a finite probability
of the holes going on the V$_2$ sites and the hopping V$_1$-V$_2$ 
would then be possible. In any case, the situation of localized 
holes on the V$_1$ sites could be recovered in the $t=0$ limit. The 
inclusion of the hopping term in our 
model makes it more general and eventually its properties could be
relevant for other compounds sharing similar structural features
with VO$_x$. In our zero temperature study, the inclusion of this
hopping is also a mimic of the activated transport that takes place 
in the real material. 

Model (\ref{hamilt}) was studied in $2\times L$ clusters by exact 
diagonalization (ED), using the Lanczos algorithm, with periodic
boundary conditions
(PBC) and by density matrix renormalization group (DMRG)\cite{dmrgrev}
with open boundary conditions (OBC). Using ED we were able to compute
static and dynamical properties on the $2\times 10$ cluster, and
using DMRG we studied $2\times 20$ and $2\times 40$ clusters although
in this case only static properties were considered. In the results
reported below obtained using the DMRG method we have retained 300-400
states in the truncation procedure.

To compute the total spin $S$ we have adopted two procedures, (i)
compute directly ${\bf S\cdot S} = S(S+1)/2$, at $S^z=0$ and (ii)
compute
$S$ as the maximum value of $S^z$ at which the ground state energy
is recovered, which implies diagonalizing the Hamiltonian in several 
subspaces of fixed $S^z$.

Spin-spin correlations $\langle S^z_j S^z_0 \rangle$ are computed 
exactly with the Lanczos algorithm in the ground state, in the
$S^z=0$ subspace. The static 
magnetic structure factor $\chi({\bf q})$ is the Fourier transform
$\chi({\bf q})=1/N \sum_{i,j} \langle S^z_j S^z_i \rangle
\exp{(i {\bf q \cdot (r_j-r_i)})}$, ($N=2 L$) where ${\bf r_j}$
(${\bf q}$) are the real space (reciprocal) vectors of a rectangular
ladder of $L\times 2$ sites where the lattice of
Fig.~\ref{fig1}, assuming equidistant sites, can be embedded.
We have verified that the results reported below do not depend
qualitatively of geometrical details (relative spatial position of
the ions) of Fig.~\ref{fig1}.
Within DMRG, we measured spin-spin correlations in the $S^z=0$ 
subspace from one of the two central sites of the cluster, one
belonging to the t-J chain and the other to the spin chain. In this
case the magnetic structure factor is computed as
$\chi({\bf q})= \sum_j \langle S^z_j S^z_0
\rangle \exp{(i {\bf q \cdot j})}$, where ``0" is one of the central
sites on the spin chain and $j$ belongs to the spin chain. In order
to compare with results for the FKL model, in most of the results
reported below for this model we have restricted the
sum to sites on the spin chain, but they are qualitatively similar
to the ones obtained with the full Fourier transform involving all
sites.

A measure of the magnetic order {\em on each chain} is provided by
the sum of the correlations:
\begin{eqnarray}
S_H= \sum_i \langle S^z_{s,i} S^z_{s,0} \rangle
\label{sum_spin}
\end{eqnarray}
where the sum extends over the sites of the spin (V$_3$-V$_2$) chain 
and $S^z_{s,0}$ is the spin of a V$_3$ site on this chain. $S_H=0$ 
corresponds to an AF ordering and $S_H>0$ to a FM or a ferrimagnetic
ordering of the spin chain.

An important property related to transport is the real part of the
optical conductivity,
$\sigma(\omega)=D \delta(\omega)+ \sigma^{reg}(\omega)$, where 
$\sigma^{reg}(\omega)$ is defined as the spectral function:
\begin{eqnarray}
\sigma^{reg}(\omega)=\frac{\pi}{\omega}\sum_n |\langle 
\psi_n| {\hat j} |\psi_0\rangle|^2 \delta(\omega - (E_n -E_0))
\label{conduct}
\end{eqnarray}
where $|\psi_n\rangle$ are eigenvectors of the Hamiltonian
(\ref{hamilt}) with energies $E_n$, and 
\begin{eqnarray}
{\hat j} = i t \sum_i ({\tilde c}^{\dagger}_{i+1 \sigma}
{\tilde c}_{i \sigma} - H.c. )
\nonumber
\end{eqnarray}
is the current operator (in units of $e=1$) and the sum extends
over the $t-J$ chain in (\ref{hamilt}). $\sigma^{reg}(\omega)$ was
computed using the continued fraction formalism.\cite{elbiorev}

From the optical conductivity, the Drude weight can be computed
as:\cite{fyescalapino,elbiorev}
\begin{eqnarray}
\frac{D}{\pi}= \frac{1}{N} (E_K - \frac{2}{\pi} \int
\sigma^{reg}(\omega) d\omega)
\label{drude}
\end{eqnarray}
\noindent
where $E_K$ is the kinetic energy, i.e., the negative ground state
average of the hopping term of the Hamiltonian (first term in
(\ref{hamilt})). $E_K$ is expressed in units of $K$, and $D$ in
units of $e^2 K$.
As usual in the continued fraction calculation of spectral functions
a Lorentzian broadening of the discrete peaks was adopted. In all
the results shown below we took a Lorentzian width $\delta=0.1$.

As mentioned above, when $J_1=J_2=0$, model (\ref{hamilt}) reduces
to uncoupled $t-J$ (where $K$ plays the role of $J$) and Heisenberg
chains. We are also interested in comparing our model with the 
FKL model with spin-$1/2$ localized spins
defined as:
\begin{eqnarray}
{\cal H} = &-& t \sum_{i,\sigma} (c^{\dagger}_{i+1 \sigma}
c_{i \sigma} + H.c. ) - J_H \sum_{i} {\bf s}_{i} \cdot
{\bf S}_{i} \nonumber \\
\label{fklm}
\end{eqnarray}
\noindent
where $J_H > 0$ is the Hund coupling, ${\bf S}_{i}$ are the
localized spins and ${\bf s}_{i}$ are the spins of the conduction
electrons. In this case the filling $n$ is defined as the filling of
the conduction chain. Notice that the t$_{2g}$ electrons in manganites
are usually modeled with spin-3/2 operators.\cite{rierahallberg} 
The frustrated t-J model is perhaps more resemblant to an 
extended version of model (\ref{fklm}) which includes an AF 
Heisenberg exchange between localized
spins.\cite{rierahallberg,yunoki,garcia1}
Model (\ref{hamilt}) would also be closer to a version of the 
FKL model which includes a Hubbard repulsion $U$ on the conduction
orbitals,\cite{schork,heldvollha} since the $t-J$ Hamiltonian appears
at an effective level when $U\gg t$.\cite{fklm_nodocc} However,
we would like to emphasize that the proposed FTJ model was not
obtained from the $U\gg t$ limit of an underlying
Hubbard-like model. In fact, most of the results reported below
belong to the case of $t \leq K$.

All the results reported below correspond to the model (\ref{hamilt})
except otherwise stated.

\section{Magnetic properties}
\label{magnetic}

We first examine the magnetic properties of this model, in particular
the presence of a ferromagnetic phase, as detected experimentally in
Ref.~\onlinecite{vanadato_tube}. The suggested mechanism that causes
this FM phase works as follows. At half-filling, due to the
frustrated AF interactions, the system would be an antiferromagnet
since it reduces to relatively isolated spin chains (V$_1$-V$_2$ 
chains in the real materials), or to 
a gapped spin liquid if these chains are dimerized. Upon electron or
hole doping, V$_1$ sites become non-magnetic, the frustration in the
V$_1$-V$_2$-V$_3$ triangles disappears, and the AF order in the
V$_2$-V$_3$ subsystem of Fig.~\ref{fig1} implies an excess of the 
$z$-component of
the total spin along one direction thus leading to ferromagnetism
or more properly, to a ferrimagnetic order.

\begin{figure}
\includegraphics[width=0.42\textwidth]{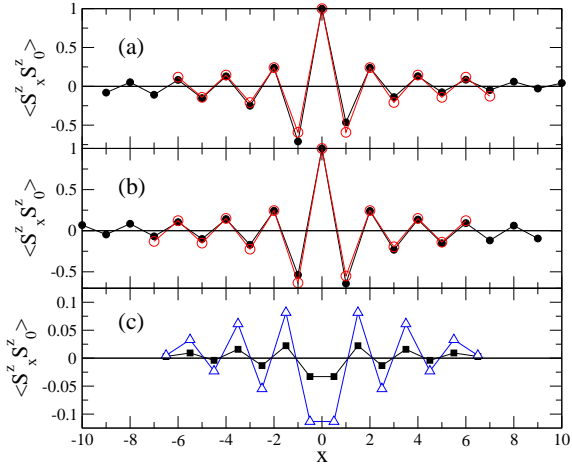}
\caption{(Color online) Spin-spin correlations 
$\langle S^z_j S^z_0 \rangle$ for the undoped system, (a) along the
V$_1$-V$_2$ chain; (b) along the V$_3$-V$_2$ chain (V$_2$ reference
site). Results were obtained for the $2\times 14$ cluster with PBC
(open circles) and for the $2\times 20$ cluster with OBC (filled
circles), $J/K=0.3$. (c) Spin-spin correlations between a V$_3$ site
and sites on the V$_1$-V$_2$ chain, $2\times 14$ cluster with PBC,
$J/K=0.3$ (squares), $J/K=1.0$ (triangles). The normalization
$\langle S^z_0 S^z_0 \rangle$=1 was adopted.}
\label{fig2}
\end{figure}

This argument is expressed in the atomic limit. For non-zero values of
$t$, there is a competition between the onsite energy, which tries to
localize holes, and the kinetic term which tries to delocalize them.
In addition, as in any
$t-J$-like model there is a competition between kinetic and
magnetic energies.

Let us first examine the undoped system. Notice that by adding
interactions between the V$_2$ sites in the bottom chain of 
Fig.~\ref{fig1}, and for the case $J=K$, one recovers the frustrated
AF Heisenberg ladder which has been extensively 
studied.\cite{hakobyan} The ground state of this model is a singlet
with a spin gap. If $J<K$, then we have an asymmetric frustrated ladder
where we can expect a similar behavior. A singlet ground state could
also be expected for our frustrated model. Some handwaving arguments
can be put forward to support this in the case $J<K$. In this case,
an AF order on the upper chain of Fig.~\ref{fig1} would dominate each
frustrated triangle and the effective interactions with couplings
$J_1$ and $J_2$ would be small. Then, both chains are essentially
uncoupled and also the lower chain would present the typical AF
ordering of Heisenberg chains. Our numerical calculations support
this picture. In Fig.~\ref{fig2} we show that for both clusters with
PBC and OBC, $J/K=0.3$, the upper and lower chains have the spin-spin 
correlations $\langle S^z_j S^z_0 \rangle$ corresponding to an AF
order (Fig.~\ref{fig2}(a) and (b) respectively). In Fig.~\ref{fig2}(c)
we show that $\langle S^z_j S^z_0 \rangle$ between a $V_3$ site and
sites of the upper chain are indeed relatively small. In fact, we
have also determined that this behavior extends up to $J/K=1.0$, as
shown in Fig.~\ref{fig2}(c). Notice the good agreement between results
obtained for PBC (with exact diagonalization) and for OBC (with DMRG)
where the reference site is one of the two central sites. It is
worth to note that the previous argument would also imply that the
undoped system is spin gapless and our numerical calculations seem
to agree with this possibility.

\begin{figure}
\includegraphics[width=0.42\textwidth]{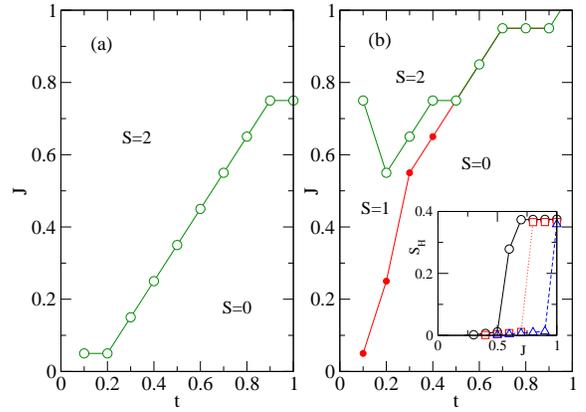}
\caption{(Color online) Phase diagram of model (\ref{hamilt}) on the
$2\times 10$ cluster with 4 holes, (a) $\epsilon_1=-2$,
(b) $\epsilon_1=-1$. The inset shows $S_H$ (defined by 
Eq.~(\ref{sum_spin})) for $t=0.3$ (circles), 0.5 (squares) and 0.7
(triangles) as a function of $J$.}
\label{fig3}
\end{figure}

In Fig.~\ref{fig3} the phase diagram of model (\ref{hamilt}) in the
plane $\{t,J\}$ is shown for the $2\times 10$ cluster and 4 holes
(doping fraction $x=0.4$). For $\epsilon_1=-2$ (Fig.~\ref{fig3}(a)),
there is an abrupt crossover from the singlet state ($S=0$) to the
ferrimagnetic state with the maximum $S$ expected at this doping,
$S=2$. As $t$ 
is increased the holes tend to be less localized on V$_1$ sites,
and then the crossover takes place at larger values of $J$. 
When $|\epsilon_1|$ is reduced, hole localization becomes less
favourable, and the crossover takes place at even larger values of
$J$ for a given $t$, as can be seen in Fig.~\ref{fig3}(b), which
corresponds to $\epsilon_1=-1$. More interesting is the fact that, for
small values of $t$, $t < 0.3$, there is an intermediate region with
$S=1$ between the $S=0$ and $S=2$ regions. The behavior of $S_H$
for several values of $t$ as a function of $J$ is shown in the inset
of Fig.~\ref{fig3}. This quantity presents a jump at the $S=0/S=2$
crossover, and since the nearest neighbor correlation 
$\langle S^z_{s,1} S^z_{s,0} \rangle < 0$, then there is a 
{\em ferrimagnetic} ordering of the spin (V$_2$-V$_3$) chain.
That is, a snapshot of this system would show an alignment of the
spins on the spin chain and at the same time an alignment of the
spins of the t-J chain with opposite orientation to the previous
one.

\begin{figure}
\includegraphics[width=0.43\textwidth]{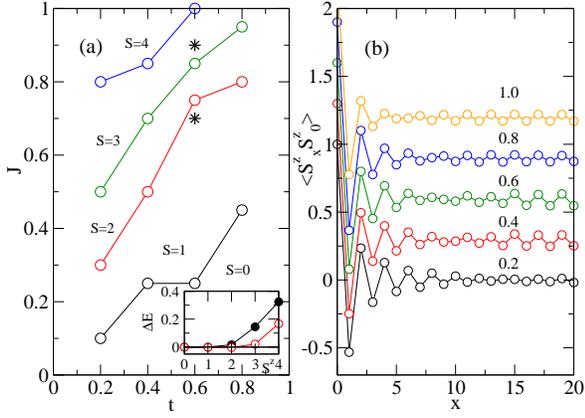}
\caption{(Color online) (a) Schematic phase diagram of model
(\ref{hamilt}) on the $2\times 20$ cluster with 8 holes, 
$\epsilon_1=-1$. At $t=0.6$, $J=0.9$ and 0.7 (points indicated with
stars), $S=8$ and 7 respectively on the $2\times 40$ cluster with 16
holes, $\epsilon_1=-1$. The inset shows $\Delta E$ (defined in the
text) vs. Sz for $t=0.2$, $J=0.2$ (filled symbols) and $J=0.4$
(open symbols), $2\times 20$ cluster. (b) Spin-spin
correlations on the $2\times 40$ cluster with 16  holes,
$\epsilon_1=-1$, $t=0.6$ and values of $J$ indicated on the
plot. The normalization $\langle S^z_0 S^z_0 \rangle$=1 was
adopted. The curves have been shifted for the sake of
clarity.}
\label{fig4}
\vspace{0.5cm}
\end{figure}

As the cluster size is increased, the intervening regions with
$0 < S < S_{max}$ develop a ``staircase", as it can be seen in 
Fig.~\ref{fig4}. In Fig.~\ref{fig4}(a), the phase diagram
in the plane $\{t,J\}$ for the $2\times 20$ cluster, with the
same doping as before, $x=0.4$, is shown. These results were
obtained with DMRG and the clusters have OBC along both directions.
The crossovers in the phase diagram of the $2\times 20$ cluster were
determined by comparing the energies obtained for different $S^z$.
In this model, as well as in the FKL model discussed below, DMRG has
a slow convergence and there is a tendency for low $S^z$ subspaces
to fall into metastable states. Then, in order to determine the
value of $S$ it is important to go to large $S^z$ and then to
extrapolate the difference in energy, $\Delta E= E(S^z)-E(S^z=0)$,
to zero. This procedure is illustrated in the inset in
Fig.~\ref{fig4}(a). In this way we have also
computed the total spin for some parameter sets, indicated in
Fig.~\ref{fig4}(a), in the $2\times 40$ cluster. These values of
$S$ for the $2\times 20$ and $2\times 40$ clusters, together with
the previous one for the $2\times 10$ cluster, seem to suggest
that the ferrimagnetic phase will survive in the thermodynamic
limit. It is interesting to note that we have also observed this
staircase in exact diagonalization studies of the FKL model on
the $L=10$ cluster (see below).
In the suggested mechanism for ferrimagnetism, the magnetic
order in the spin chain would still be an AF one. It can be seen
in Fig.~\ref{fig4}(b) that the spin-spin correlations 
$\langle S^z_j S^z_0 \rangle$ for the $2\times 40$ clusters,
$x=0.4$, are indeed essentially AF except for the presence of a small
kink.  The same behavior was found for the $2\times 10$ cluster
with PBC and the same doping in the $S=2$ region.

\begin{figure}
\includegraphics[width=0.43\textwidth]{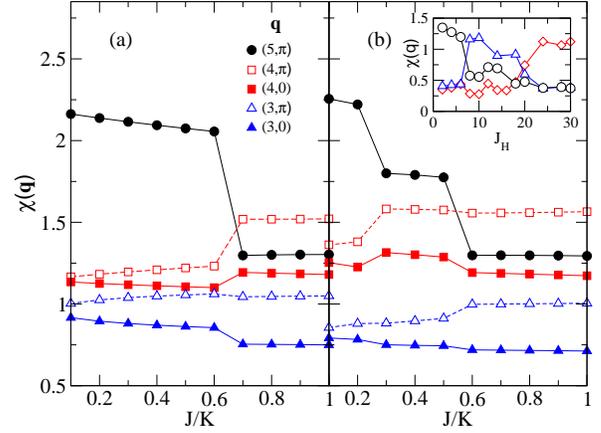}
\caption{(Color online) Static magnetic structure factor $\chi(\bf q)$
as a function of $J$ along cuts of constant $t$ on the phase diagrams
of Fig.~\ref{fig3}, $2\times 10$ cluster, 4 holes, (a) $\epsilon_1=-2$,
$t=0.8$, (b) $\epsilon_1=-1$, $t=0.2$. The values of $\bf q$ are
indicated on the plot, $q_x$ in units of $\pi/5$. The inset
shows $\chi(q)$ for the ferromagnetic Kondo lattice model as a
function of $J_H$, $L=10$, $n=0.6$, $q_x=3$ (circles), 2 (triangles)
and 1 (diamonds) in units of $\pi/5$.}
\label{fig5}
\vspace{0.5cm}
\end{figure}

In order to study more systematically the behavior of the spin-spin
correlations in the regions of different total spin let us turn to
study the static magnetic structure factor, $\chi(\bf q)$. Results
for $\chi(\bf q)$ in the $S^z=0$ subspace along a ``cut" at $t=0.8$
in Fig.~\ref{fig3}(a), and at $t=0.2$ in Fig.~\ref{fig3}(b) are
depicted in Fig.~\ref{fig5}(a) and (b) respectively. These results
were obtained by exact diagonalization and the first definition of
$\chi(\bf q)$ discussed in Section \ref{model} was adopted. In both
cases considered, in the $S=0$ region (small values of $J/K$),
$\chi(\bf q)$ is maximum at $q_x=\pi$ ($x$ is along the chain
direction), indicating an AF order, irrespective of $q_y=0$ and
$\pi$. In Fig.~\ref{fig5}(a), in the $S=2$ region ($J/K \geq 0.7$),
the peak of $\chi(\bf q)$ shifts away from $q_x=\pi$, indicating an
incommensurate (IC) order. In Fig.~\ref{fig5}(b), $\chi(\pi,0)$ has a
sudden drop upon entering in the $S=1$ region ($J/K \sim 0.3$) while
$\chi(3\pi/5,0)$ and $\chi(4\pi/5,0)$ increase. A similar behavior
occurs when entering in the $S=2$ region ($J/K \geq 0.6$) but now
the maximum of $\chi(\bf q)$ is located at ${\bf q}=(4\pi/5,0)$ 
corresponding to the IC order. This behavior do not depend
qualitatively on other choices of the set of reciprocal vectors
entering in the Fourier transform. The inset shows $\chi(q)$ for
localized spins for the FKL model as a function of $J_H$ for the
same chain length, $L=10$, and filling $n=0.6$. In this case,
$q$ is reduced from $3 \pi/5$ to its minimum possible nonzero
value, $\pi/5$, which corresponds to FM order in the $S^z=0$
subspace with PBC. It is clear the presence of two crossovers, at
$J_H\sim 6.5$ and at $\sim 19.5$, where also the total spin of the
system jumps from 0 to 1 and then from 1 to 2, respectively.

\begin{figure}
\includegraphics[width=0.43\textwidth]{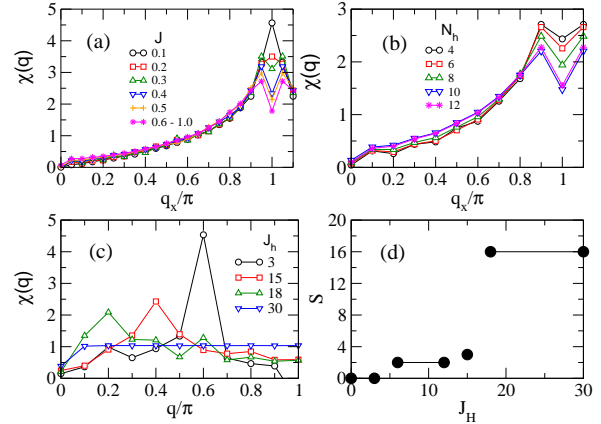}
\caption{(Color online) Static magnetic structure factor $\chi(q)$
computed with DMRG (a) in the $2\times 40$ cluster, for several
values of $J$ indicated on the plot, $\epsilon_1=-1$, and $t=0.3$;
(b) in the $2\times 20$ cluster for $\epsilon_1=-1$, $t=0.4$, $J=1$
at several hole doping $N_h$ indicated on the plot; (c) for the FKL
model in the $L=20$ chain, $x=0.6$, for several $J_H$ indicated on 
the plot. (d) Total spin of the ground state of the FKL model as a
function of $J_H$, $L=20$, $x=0.6$.}
\label{fig6}
\end{figure}

Similar behavior of the static magnetic structure factor is observed
in the $2\times 20$ and $2\times 40$ clusters with OBC.
Fig.~\ref{fig6}(a) shows $\chi(q)$ in the $2\times 40$ as a function
of $q_x$ for various values of $J$, at $t=0.3$, obtained with DMRG. In
this case only the spins of the spin chain have been included in the
Fourier transform but results involving all spin-spin correlations
are similar.  It can be seen that, for a fixed $t$, $\chi(q)$ remains
approximately unchanged during some intervals of $J$. Each of
these intervals of $J$ can be related to a given value of the total
spin $S$, but there is not necessarily a one-to-one correspondance.
As hole doping increases,
the IC peak of $\chi(q)$ becomes sharper but it remains at
${\bf q}=((L\pm 1)\pi/L,0)$, as it can be seen in Fig.~\ref{fig6}(b).
We will term these states with $S>0$ as IC ferrimagnetic states.
This behavior, which was also observed for the other clusters
studied, is different from the
one reported for the 1D FKL model.\cite{dagotto} In the FKL model,
when the electron occupation is reduced from half-filling at a fixed
value of the Hund coupling, the peak of the static magnetic
structure factor, located at $2k_F=n\pi$, is reduced from its AF
value $q_x=\pi$ to $q_x= 0$ in the FM region. In Fig.~\ref{fig6}(c)
we show for comparison the evolution of $\chi(q)$ for the FKL model as
$J_H$ is increased at an equivalent cluster size and filling. In 
contrast with Fig.~\ref{fig6}(a), the peak of $\chi(q)$ moves away
from its IC position as it approaches the crossover to the FM phase
located at $J_H \sim 18$. Inside this phase, $\chi(q)$ presents the
typical form of a FM order in the $S^z=0$ subspace. In 
Fig.~\ref{fig6}(d) we show the corresponding evolution of the total 
spin of the ground state. Although DMRG calculations have a very slow
convergence for this model, a problem that is also present in Lanczos 
diagonalization, there are apparently regions in $J_H$ with
intermediate values of $S$ which may correspond to the various 
positions of the peak of $\chi(q)$ shown in Fig.~\ref{fig6}(c).
However, this staircase in the values of $S$ is not as pronounced as
the one for model (\ref{hamilt}). Moreover, these results could be 
affected by finite size effects.  These 
differences between the two models illustrate the IC ferrimagnetic
character of the ground state in the present model
as opposed to the true ferromagnetic state present in the FKL model.

\section{Transport properties}
\label{transport}

For the possible applications of VO$_x$ in spintronics devices, it
is essential to determine the correlation between its magnetic 
and transport properties. That is, one would expect that, as in
manganites, transport is enhanced as the total spin of the ground
state increases.

\begin{figure}
\includegraphics[width=0.42\textwidth]{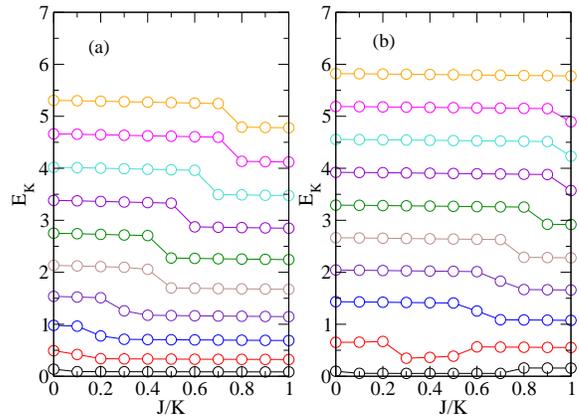}
\caption{(Color online) Kinetic energy (in units of $K$) as a 
function of $J$ on the $2\times 10$ cluster with 4 holes, 
$t=0.1, 0.2, \ldots, 1.0$ from bottom to top (a)  $\epsilon_1=-2$,
(b) $\epsilon_1=-1$.}
\label{fig7}
\end{figure}

An elementary indication of charge mobility is provided by the
kinetic energy, $E_K$.  Fig.~\ref{fig7}(a) and (b) show $E_K$
obtained by ED for the $2\times 10$ cluster with
4 holes in the $S^z=0$ sector for $\epsilon_1=-2$ and $\epsilon_1=-1$
respectively. It can be seen in Fig.~\ref{fig7} that as $J$ is
increased for a fixed $t$ there are step-like reductions of
$E_K$ at values of $J$ which exactly coincide with the boundaries
between regions with increasing values of $S$ in the phase diagrams
shown in Fig.~\ref{fig3}. That is, the increase of $S$ is
accompanied by an increased hole localization. Although it is not
perceptible in the scale of Fig.~\ref{fig7}, in addition to these
steps, $E_K$ slowly decreases as $J$ is increased due to the 
competing magnetic energy. 

For small values of the hopping constant $t$ ($t \leq 0.2$), small
{\em increases} of $E_K$ as a function of $J$ can also be
observed in Fig.~\ref{fig7}(b). This behavior can be understood
from the fact that in the regime $t << K$, in order to gain magnetic
energy, the spins on the t-J chain have the tendency to form
hole-free islands thus leading
to a phase separated state. The exchange couplings $J_1$ and $J_2$
connecting the $t-J$ chain with the Heisenberg chain are 
competing with $K$ and then opposite to the formation of those
islands. Then, as $J_1=J_2=J$ increase there will be at some
point a breaking of that phase separated state thus favoring hole
delocalization and an increase in kinetic energy.

\begin{figure}
\includegraphics[width=0.42\textwidth]{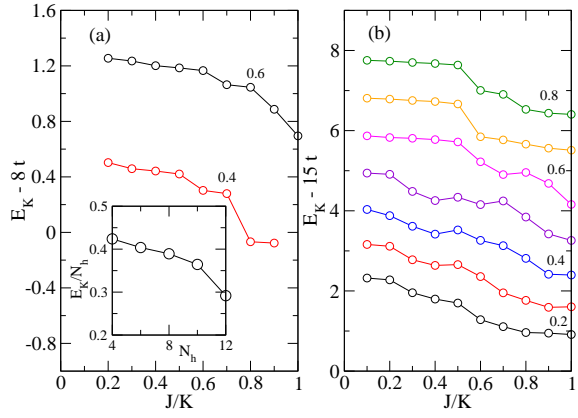}
\caption{(Color online) Kinetic energy $E_K$ (in units of $K$) as a
function of $J$, for several values of $t$ indicated on the plot,
$\epsilon_1=-1$, (a) for the $2\times 20$ cluster with 8 holes, (b) for
the $2\times 40$ cluster with 16 holes. $E_K$ has been shifted as 
indicated in the label of the $y$ axis for clarity. The inset shows
$E_K$ per hole in the $2\times 20$ cluster as a function of the number
of holes.}
\label{fig8}
\vspace{0.5cm}
\end{figure}

A similar behavior of the kinetic energy has been observed in
the $2\times 20$ cluster with 8 holes, and for the $2\times 40$
cluster with 16 holes as shown in Fig.~\ref{fig8}(a) and (b)
respectively. In order to clearly appreciate the evolution of
$E_K$ for various values of $t$, the values of $E_K$ have been
shifted by multiples of $t$ as indicated in the vertical axis.
For a given value of $t$ there is a general
decrease of $E_K$ as $J$ increases, with some sudden drops which
coincide with the successive increases of $S$ shown in the phase
diagram of Fig.~\ref{fig4}. It can be seen in the inset of 
Fig.~\ref{fig8} that the kinetic energy per doped hole also
decreases as hole doping is increased. In the  FKL model,
$E_K$ also decreases as $J_H$ increases at a given
density\cite{rierahallberg} due again to the competition with 
the magnetic energy (see further discussion below) but $E_K$
{\em per electron} increases for a fixed $J_H$.

\begin{figure}
\includegraphics[width=0.42\textwidth]{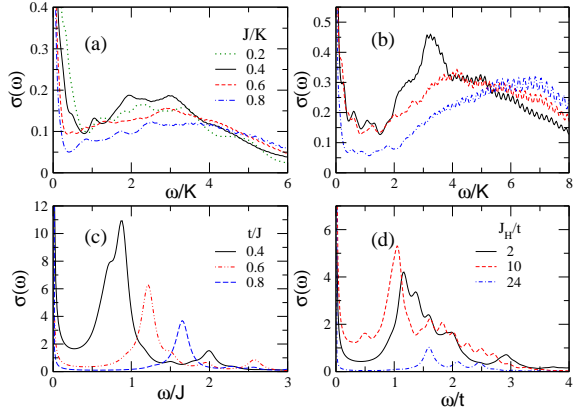}
\caption{(Color online) Optical conductivity (in units of $e^2$) for 
the $2\times 10$ cluster with 4 holes, $\epsilon_1=-1$, (a) $t=0.4$,
(b) $t=0.9$, and for several values of $J$ indicated in the plot.
(c) Optical conductivity for the 1D extended $t-J$ model on the
10-site chain, $J=1$, $\epsilon_1=-1$, 4 holes, and several values of
$t/J$ as indicated in the plot. (d) Optical conductivity for the FKL
model (Eq.~\ref{fklm}) on the 10-site chain, 6 conduction electrons, 
and several values of $J_H/t$ indicated on the plot.
}
\label{fig9}
\end{figure}

Fig.~\ref{fig9} shows the regular part of the optical conductivity 
$\sigma^{reg}(\omega)$ computed on the $2\times 10$ cluster with 4
holes, (a) $t=0.4$, (b) $t=0.9$, $\epsilon_1=-1$. Excluding the peak
at $\omega=0$ which should be replaced by the Drude
weight,\cite{elbiorev} $\sigma^{reg}(\omega)$ and $\sigma(\omega)$
coincide, so we will use $\sigma(\omega)$ in the following. For
both values of $t$, $\sigma(\omega)$ presents a broad maximum starting
at $\sim 4t$. For a fixed $t$, this maximum first increases 
as $J$ is increased from 0.2 to 0.4, and then it decreases
as $J$ is further increased. The position of this maximum
is shifted to higher values of $\omega$.  In Fig.~\ref{fig9}(c),
$\sigma(\omega)$ is shown for the 1D $t-J$ model, on the
10-site chain with 4 holes. Notice that $J$ is actually $K$ in the
notation of the Hamiltonian (\ref{hamilt}) and hence we adopted the
value $J=1$. Also, for the sake of comparison, an on-site potential
every two sites, with $\epsilon_1=-1$, was added to the usual $t-J$
Hamiltonian. The behavior of $\sigma(\omega)$ in both
models is clearly different. In the extended $t-J$ model there
is a peak located at $\omega\sim 2t$ instead of the 
broad feature observed for the FTJ model. In the
present model, $\sigma(\omega)$ is more resemblant to the behavior
in two-dimensional clusters.\cite{elbiorev} In two dimensions,
the competition between kinetic and magnetic energies is more
important than in one dimension, where in general spin-charge
separation holds. $\sigma(\omega)$ for the FTJ model 
is also more similar to the one for the FKL model (\ref{fklm}), as
shown in Fig.~\ref{fig9}(d) for an equivalent chain size and 
filling: the incoherent structure is broad, and the spectral weight
below this region (but at finite $\omega$) may correspond in
both models to the presence of a pseudogap. However, in the FKL
model there is a pronounced change of $\sigma(\omega)$ when 
entering in the FM region ($J_H/t \geq 19$) implying an important
transfer of weight from the incoherent part to the Drude peak.
This behavior in the FM region is consistent with that reported 
in previous studies,\cite{yunoki,heldvollha,fklm_nodocc} and it
has been understood within the DE model.\cite{furukawa}
The small remaining incoherent part does not scale with $J_H/t$
and it is almost certainly a finite size effect.
This suppression of the incoherent part of $\sigma(\omega)$
is virtually absent in the frustrated t-J model for VO$x$.
Notice also the much smaller scale of $\sigma(\omega)$ for 
the FTJ model with respect to those for the extended $t-J$
and FKL models. These different scales are in part due to the fact
that $t$ was used as the scale of energy in the latter models
while $K$ was used in our model. However, even after correcting for 
this factor,
there is still an order of magnitude between $\sigma^{reg}(\omega)$
in the FTJ model and the other two models.

\vspace{0.4cm}
\begin{figure}
\includegraphics[width=0.42\textwidth]{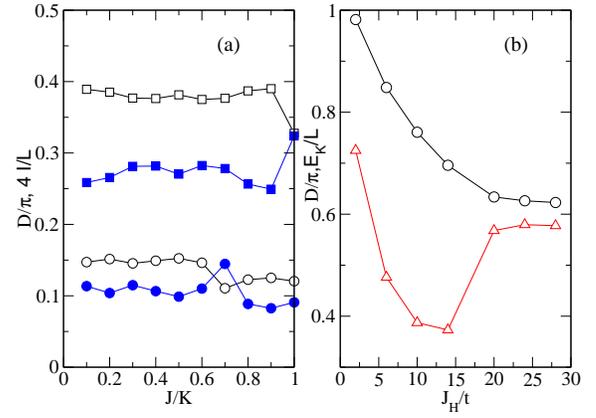}
\caption{(Color online) (a) Drude weight (open symbols) and
the integral of $\sigma^{reg}(\omega)$ (full symbols) for model 
(\ref{hamilt}) for the $2\times 10$ cluster with 4 holes,
$\epsilon_1=-1$, as a function of $J$, $t=0.4$ (circles) and 
$t=0.9$ (squares).  (b) Drude weight (triangles) and kinetic
energy per site (circles) for the FKL model for $L=10$ cluster, 6 
conduction electrons, as a function of $J_H/t$.}
\label{fig10}
\end{figure}

Finally, the Drude weight for the $2\times 10$ cluster with 4 holes,
$\epsilon_1=-1$, is shown in Fig.~\ref{fig10}(a) as a function
of $J$, for $t=0.4$ and $t=0.9$. It is instructive to examine first the
behavior of the integral $I$ of $\sigma^{reg}(\omega)$ (second term in 
Eq.~(\ref{drude})) which measures the weight of the incoherent part of
the optical conductivity. This quantity, for a given value of $t$,
presents a maximum just at the crossover between the $S=0$ and $S >0$
phases, and then it smoothly decreases upon further increase of $J$.
However, this integral is much smaller than the kinetic energy
(notice the factor of 4 in Fig.~\ref{fig10}). As a result, as it can
be clearly seen in this Figure, the Drude weight
qualitatively follows the behavior of the kinetic energy, shown
in Fig. ~\ref{fig7}(b). That is, for a given value of $t$ it presents 
sudden drops at the values of $J$ at which the total spin of the
ground state increases.
The behavior of the Drude weight of model (\ref{hamilt}) is quite
different from that of the FKL model. In Fig.~\ref{fig10}(b)
we show the Drude weight for the later model on the $L=10$ chain
with PBC, 6 conduction electrons, obtained by ED.
Although the kinetic energy monotonically decreases when $J_H/t$
increases due to the competition with the magnetic 
energy\cite{rierahallberg}, the Drude weight first decreases while
still inside
the IC phase and then it is strongly enhanced upon entering in the
FM phase, approximately at $J_H/t\sim 19$, as mentioned above. This
sudden increase of the Drude weight is consistent with the 
noticeable suppression of the incoherent structure of the optical
conductivity in the FM region, observed in Fig.~\ref{fig9}(d).

\section{Conclusions}
\label{conclusions}

We have proposed a 1D model to describe a mechanism suggested 
to explain the appearance of a FM phase in vanadate nanotubes upon
doping. This model consists of a t-J chain coupled to a Heisenberg
chain with frustrated exchange interactions.

Magnetic properties show the presence of an ascending ``staircase" of 
the total spin $S$ as the magnetic couplings $J_1=J_2=J$ increase for
a fixed value of the hopping integral $t$. These phases with $S >0$ 
have a ferrimagnetic origin and spin-spin correlations do not
present the typical behavior of a ferromagnetic state but they
correspond to an incommensurate phase which is described by a peak of 
the structure magnetic factor located at ${\bf q}=((L\pm 1)\pi/L,0)$
independently of the doping fraction. Our results suggest
that this ferrimagnetic phase will survive in the thermodynamic
limit. We have also observed this staircase in $S$ in the FKL model, 
a result which was not reported to our knowledge 
in previous studies on this model.\cite{dagotto,garcia2,kienert}
However, we cannot exclude at this point the possibility of this 
result being a mere finite size effect.
In this model, we have also observed that the magnetic peak moves
from its IC value to its FM value in a step-like fashion as 
$J_H/t$ increases {\em at a fixed density}.

The optical conductivity for the 1D model for VO$_x$ shows a broad
incoherent structure with indications of the presence of a pseudogap,
and spectral weight that is transferred to higher frequencies as 
$J$ is increased for a given value of the hopping constant $t$.
This behavior is different from the one observed in isolated
$t-J$ chains, and more relevant to the topic of ferromagnetic
metals, it is also very different from the one observed in the 1D
FKL model. This different behavior in $\sigma(\omega)$ is 
translated to the Drude weight. In the FKL model there is
a strong enhancement of the Drude weight upon entering in the
FM phase, while in the present model the Drude weight essentially
follows the behavior of the kinetic energy.

In summary, in the proposed model for VO$_x$, at least in its 
1D version, there is not a true ferromagnetic order but rather an
IC ferrimagnetic one, and contrary to what happens in the FKL
model, electronic transport is somewhat suppressed by this
ferrimagnetic order. More realistic variants of the model here
studied would include a three-chain ladder or a two-dimensional
lattice, a Hubbard interaction instead of the t-J one, and a
dimerization on the t-J chain. By increasing the dimensionality
it could be possible to achieve a ratio between V$_1$, V$_2$
and V$_3$ ions closer to the real system. It would be closer to
the actual materials to include a direct hopping between V$_2$
sites.  Even for the present model, we
have only explored a small region in the parameter space, and in
particular it seems promising to examine the case $J_1 \neq J_2$,
as we plan to perform in future works.

\acknowledgments
We thanks A. Dobry, C. Gazza, G. B. Martins, M. E. Torio and S.
Yunoki for useful discussions.


\begin{thebibliography}{}

\bibitem{wolf} S. A. Wolf, D. D. Awschalom, R. A. Buhrman, J. M. Daughton,
       S. von Moln\'ar, M. L. Roukes, A. Y. Chtchelkanova, and D. M.
       Treger, Science {\bf 294}, 1488 (2001).
       S. Maekawa and T. Shinjo, {\it Spin Dependent Transport in Magnetic
       Nanostructures}. (Taylor \& Francis, London, 2002).

\bibitem{lamno} M. B. Salamon and M. Jaime, Rev. Mod. Phys {\bf 73}, 
        583 (2001);
        J. M. D. Coey, M. Viret, and S. von Molnar, Adv. Phys. 
        {\bf 46}, 7268 (1992); T. Kaplan and S. Mahanti, (eds.),
        {\it Physics of Manganites}, (Kluwer Academic/Plenum Publishers,
        New York, 1999).  

\bibitem{colossal} A. P. Ramirez, J. Phys.: Condens. Matter {\bf 9},
        8171 (1997), and references therein. 

\bibitem{vanadato_tube} L. Krusin-Elbaum, D. M. Newns, H. Zeng, V.
         Derycke, J. Z. Sun, and R. Sandstrom, Nature {\bf 431}, 672
         (2004).

\bibitem{manganato_tube} P. Levy,  A. G. Leyva, H. E. Troiani, and 
       R. D. S\'anchez, Appl. Phys. Lett. {\bf 83}, 5247 (2003); J. 
       Curiale, R. D. S\'anchez, H. E. Troiani,  A. G. Leyva, and P. 
       Levy, Appl. Phys. Lett. {\bf 87}, 043113 (2005).

\bibitem{pasupathy} A. N. Pasupathy, R. C. Bialczak, J. Martinek, J. E.
        Grose, L. A. K. Donev, P. L. McEuen, and D. C. Ralph, Science
       {\bf 306}, 86 (2004).

\bibitem{martinek} J. Martinek, M. Sindel, L. Borda, J. Barnas, J.
       K\"onig, G. Sch\"on, and J. von Delft, Phys. Rev. Lett. {\bf 91},
        247202 (2003); C. J. Gazza, M. E. Torio, and J. A. Riera, Phys.
        Rev. B {\bf 73}, 193108 (2006).

\bibitem{votube2} E. Vavilova, I. Hellmann, V. Kataev, C. T\"aschner,
      B. B\"uchner, and R. Klingeler, Phys. Rev. B {\bf 73}, 144417
       (2006).

\bibitem{votube0} F. Krumeich, H.-J. Muhr, M. Niederberger, F. Bieri, 
        B. Schnyder, and R. Nesper, J. Am. Chem. Soc. {\bf 121}, 8324
       (1999).

\bibitem{votube1} X. Liu, C. T\"aschner, A. Leonhardt, M. H. R\"ummeli,
        T. Pichler, T. Gemming, B. B\"uchner, and M. Knupfer, Phys.
        Rev.B {\bf 72}, 115407  (2005).

\bibitem{carbo_nano} S. J. Tans, A. R. M. Verschueren, C. Dekker, Nature 
      {\bf 393}, 49 (1998); K. Tsukagoshi, B. W. Alphenaar, H. Ago,
       Nature {\bf 401}, 572 (1999); M. J. Biercuk, S. Garaj, N. Mason,
       J. M. Chow, C. M.  Marcus, Nano Letters {\bf 5}, 1267 (2005).

\bibitem{rierahallberg} J. Riera, K. Hallberg and E. Dagotto,
         Phys. Rev. Lett. {\bf 79}, 713 (1997).

\bibitem{dagotto} E. Dagotto, S. Yunoki, A. L. Malvezzi, A. Moreo, J.
        Hu, S. Capponi, D. Poilblanc, and N. Furukawa, Phys. Rev. B
        {\bf 58}, 6414 (1998).

\bibitem{two-orbs} T. Hotta, M. Moraghebi, A. Feiguin, A. 
         Moreo, S. Yunoki, and E. Dagotto, Phys. Rev. Lett. {\bf 90}, 
         247203 (2003).

\bibitem{garcia2} D. J. Garcia, K. Hallberg, B. Alascio, and M. Avignon,
        Phys. Rev. Lett. {\bf 93}, 177204 (2004).

\bibitem{kienert} J. Kienert and W. Nolting, Phys. Rev. B {\bf 73},
         224405 (2006).

\bibitem{dmrgrev} {\it Density-Matrix Renormalization}, edited by I.
        Peschel, X. Wang, M. Kaulke, and K. Hallberg,
         (Springer, Berlin, 1999); U. Schollw\"{o}ck, Rev.
         Mod. Phys. 77, 259 (2005). 

\bibitem{elbiorev} E. Dagotto, Rev. Mod. Phys. {\bf 66}, 763 (1994).

\bibitem{fyescalapino} R. M. Fye, M. J. Martins, D. J. Scalapino, J.
        Wagner, and W. Hanke Phys. Rev. B {\bf 44}, 6909 (1991).

\bibitem{yunoki} S. Yunoki and A. Moreo, Phys. Rev. B {\bf 58}, 6403
        (1998).

\bibitem{garcia1} D. J. Garcia, K. Hallberg, C. D. Batista, M. Avignon, 
        and B. Alascio,
        Phys. Rev. Lett. {\bf 85}, 3720 (2000).

\bibitem{schork} See, e.g., T. Schork, S. Blawid, and J.-i. Igarashi, 
        Phys. Rev. B {\bf 59}, 9888 (1999).

\bibitem{heldvollha} K. Held and D. Vollhardt, Phys. Rev. Lett.
        {\bf 84}, 5168 (2000).

\bibitem{fklm_nodocc} For a study of the no-double occupancy version of
        the FKL model see P. Horsch, J. Jakli\v{c}, and F. Mack,
        Phys. Rev. B {\bf 59}, 6217 (1999).

\bibitem{hakobyan} T. Hakobyan, J. H. Hetherington, and M. Roger, 
        Phys. Rev. B {\bf 63}, 144433 (2001), and references therein.

\bibitem{furukawa} N. Furukawa, J, Phys. Soc. Jpn. {\bf 64}, 3164
       (1995).

\end{thebibliography}
\end{document}